\documentclass[twocolumn,showpacs,amsmath,amssymb]{revtex4}
\usepackage{graphicx}
\usepackage{amsfonts}
\usepackage{dcolumn}
\usepackage{bm}

\usepackage{CJK}

\begin{document}
\title{Critical noise of majority-vote model on complex networks}

\author{Hanshuang Chen$^{1}$}\email{chenhshf@ahu.edu.cn}

\author{Chuansheng Shen$^2$}

\author{Gang He$^1$}

\author{Haifeng Zhang$^3$}

\author{Zhonghuai Hou$^{4}$}\email{hzhlj@ustc.edu.cn}

\affiliation{$^{1}$School of Physics and Material Science, Anhui
University, Hefei, 230039, China \\$^2$Department of Physics, Anqing
Normal University, Anqing, 246011, China \\ $^3$School of Mathematical Science, Anhui University, Hefei, 230039, China \\
$^{4}$Hefei National Laboratory for Physical Sciences at Microscales
\& Department of Chemical Physics, University of
 Science and Technology of China, Hefei, 230026, China}

\date{\today}

\begin{abstract}
The majority-vote model with noise is one of the simplest
nonequilibrium statistical model that has been extensively studied
in the context of complex networks. However, the relationship
between the critical noise where the order-disorder phase transition
takes place and the topology of the underlying networks is still
lacking. In the paper, we use the heterogeneous mean-field theory to
derive the rate equation for governing the model's dynamics that can
analytically determine the critical noise $f_c$ in the limit of
infinite network size $N\rightarrow \infty$. The result shows that
$f_c$ depends on the ratio of ${\left\langle k \right\rangle }$ to
${\left\langle {{k^{{3 \mathord{\left/ {\vphantom {3 2}} \right.
 \kern-\nulldelimiterspace} 2}}}} \right\rangle }$,
where ${\left\langle k \right\rangle }$ and ${\left\langle {{k^{{3
\mathord{\left/ {\vphantom {3 2}} \right.
 \kern-\nulldelimiterspace} 2}}}} \right\rangle }$ are the average degree and the ${{3 \mathord{\left/
{\vphantom {3 2}} \right.
 \kern-\nulldelimiterspace} 2}}$ order moment of
degree distribution, respectively. Furthermore, we consider the
finite size effect where the stochastic fluctuation should be
involved. To the end, we derive the Langevin equation and obtain the
potential of the corresponding Fokker-Planck equation. This allows
us to calculate the effective critical noise $f_c(N)$ at which the
susceptibility is maximal in finite size networks. We find that the
$f_c-f_c(N)$ decays with $N$ in a power-law way and vanishes for
$N\rightarrow \infty$. All the theoretical results are confirmed by
performing the extensive Monte Carlo simulations in random
$k$-regular networks,  Erd\"os-R\'enyi random networks and
scale-free networks.

\end{abstract}
 \maketitle

\section{Introduction}

Equilibrium and nonequilibrium phase transitions in ensembles of
complex networked systems have been a subject of intense research in
the field of statistical physics and many other disciplines
\cite{PRP06000175,PRP08000093,RMP08001275,PRP2014}. Owing to the
inherent randomness and heterogeneity in the interacting patterns,
phase transitions on complex networks are drastically different from
those on regular lattices in Euclidean space. Examples range from
the anomalous behavior of Ising model
\cite{PHA02000260,PLA02000166,PRE02016104,EPB02000191,PhysRevLett.104.218701}
to a vanishing percolation threshold
\cite{PhysRevLett.85.4626,PRL00005468} and the absence of epidemic
thresholds that separate healthy and endemic phases
\cite{PRL01003200,PhysRevLett.90.028701,PhysRevLett.111.068701} as
well as explosive emergence of phase transitions
\cite{Science323.1453,PhysRevLett.103.255701,PhysRevLett.103.168701,
PhysRevLett.103.135702,PhysRevLett.105.255701,PhysRevLett.106.225701,Science333.322,
Nature2010,PhysRevLett.105.048701,PhysRevLett.107.195701,
PhysRevLett.106.128701,PhysRevLett.108.168702,NatPhys2014}. So far,
unveiling the relationship between the onset of phase transitions
and the topology of the underlying networks is still a topic of
considerable attention.

The majority-voter (MV) model is a simple nonequilibrium Ising-like
system with up-down symmetry that presents an order-disorder phase
transition at a critical value of noise \cite{RMP09000591}. Since
Oliveira pointed out that the MV model on a square lattice belongs
to the universality class of the equilibrium Ising model
\cite{JSP1992}, the model has been extensively studied in the
context of complex networks, including random graphs
\cite{PhysRevE.71.016123,PA2008}, small world networks
\cite{PhysRevE.67.026104,IJMPC2007,PA2015}, scale-free networks
\cite{IJMPC2006(1),IJMPC2006(2)}, and some others
\cite{PhysRevE.75.061110,PhysRevE.77.051122,PhysRevE.81.011133,PA2011,PhysRevE.86.041123,PhysRevE.89.052109}.
These results showed that the critical exponents are generally
dependent on the underlying interacting substrates. However, the
studies of pervious works are mainly based on numerical simulations.
Especially, the analytical determination of the critical noise is
still lacking at present.

For this purpose, in this paper we employ the heterogeneous
mean-field theory to derive the rate equation for governing the MV
model's dynamics on undirected networks. According to linear
stability analysis, we determine the critical point of noise $f_c$,
the onset of an order-disorder phase transition in the limit of
infinite size networks $N \rightarrow \infty$. The analytical result
shows that $f_c$ is related to the ratio of the first moment to the
${3 \mathord{\left/ {\vphantom {3 2}} \right.
\kern-\nulldelimiterspace} 2}$ order moment of degree distribution.
Furthermore, we derive the Langevin equation to study the effect of
stochastic fluctuation on finite size networks. By solving the
potential of the corresponding Fokker-Planck equation, we calculate
the susceptibility as a function of noise and determine the
effective critical noise $f_c(N)$ on finite size networks at which
the susceptibility is maximal. We find that the difference
$f_c-f_c(N)$ decays in power-law ways with $N$. Extensive Monte
Carlo (MC) simulations are performed on diverse network types to
validate the theoretical results.

\section{Model}
We consider the MV model with noise on complex networks defined by a
set of spin variables $\{\sigma_i\}$ $(i = 1, \ldots ,N)$, where
each spin is associated to one node of the underlying network and
can take the values $\pm 1$. The system evolves as follows: for each
spin $i$, we first determine the majority spin of $i'$s
neighborhood. With probability $f$ the node $i$ takes the opposite
sign of the majority spin, otherwise it takes the same spin as the
majority spin. The probability $f$ is called the noise parameter and
plays a similar role of temperature in equilibrium spin systems. In
this way, the single spin flip probability can be written as
\begin{eqnarray}
w({\sigma _i}) = \frac{1}{2}\left[ {1 - \left( {1 - 2f}
\right){\sigma _i}S\left( {\sum\limits_j {{a_{ij}}{\sigma _j}} }
\right)} \right], \label{eq1}
\end{eqnarray}
where $S(x)=sgn(x)$ if $x\neq0$ and $S(0)=0$. In the latter case the
spin $\sigma_i$ is flipped to $\pm 1$ with probability equal to ${1
\mathord{\left/ {\vphantom {1 2}} \right. \kern-\nulldelimiterspace}
2}$. The elements of the adjacency matrix of the underlying network
are defined as $a_{ij}=1$ if nodes $i$ and $j$ are connected and
$a_{ij}=0$ otherwise. In the case $f=0$, the majority-vote model is
identical to the zero temperature Ising model
\cite{PNAS2005,JSM2006}.

\section{Results}
To proceed a mean-field treatment, we first define $q_k$ as the
probability that a node of degree $k$ is in $+1$ state, and $Q$ as
the probability that for any node in the network, a randomly chosen
nearest neighbor node is in $+1$ state. Furthermore, for any node
the probability that a randomly chosen nearest neighbor node has
degree $k$ is ${{kP(k)} \mathord{\left/
 {\vphantom {{kP(k)} {\left\langle k \right\rangle }}} \right.
 \kern-\nulldelimiterspace} {\left\langle k \right\rangle }}$, where $P(k)$ is degree distribution
defined as the probability that a node chosen at random has degree
$k$ and $\left\langle k \right\rangle $ is the average degree
\cite{RMP08001275}. It is supposed to be reasonable only in networks
without degree correlation. The probabilities $q_k$ and $Q$ satisfy
the relation
\begin{eqnarray}
Q = \sum\limits_k {kP(k)} {{{q_k}} \mathord{\left/
 {\vphantom {{{q_k}} {\left\langle k \right\rangle }}} \right.
 \kern-\nulldelimiterspace} {\left\langle k \right\rangle }}. \label{eq2}
\end{eqnarray}
Thus, we can write rate equations for $q_k$ as
\begin{eqnarray}
{{\dot q}_k} &=&  - {q_k}(1 - {\psi _k}) + (1 - {q_k}){\psi _k}
\nonumber \\ &=&  - {q_k} + {\psi _k}, \label{eq3}
\end{eqnarray}
where ${\psi _k}$ is is the probability that a node of degree $k$
takes the $+1$ value, which can be expressed as
\begin{eqnarray}
{\psi _k(Q)} = (1 - f){\varphi _k}(Q) + f({1-\varphi _k}(Q)).
\label{eq4}
\end{eqnarray}
Here, ${\varphi _k}(Q)$ is the probability that a node of degree $k$
with $+1$ state takes the majority rule, which can be written by a
binomial distribution,
\begin{eqnarray}
{\varphi _k}(Q) = \sum\limits_{n = \left\lceil {k/2} \right\rceil
}^k {\left(1 - \frac{1}{2}{\delta _{n,k/2}}\right) C_k^n} {Q^n}{(1 -
Q)^{k - n}}, \label{eq5}
\end{eqnarray}
where $\left\lceil \cdot \right\rceil$ is the ceiling function,
$\delta$ is the Kronecker symbol, and $C_k^n = {{k!} \mathord{\left/
 {\vphantom {{k!} {[n!(k - n)!])}}} \right.
 \kern-\nulldelimiterspace} {[n!(k - n)!]}}$ are the binomial coefficients. By introducing Eq.(\ref{eq3}) into Eq.(\ref{eq2}) we obtain a closed rate equation for
the quantity $Q$,
\begin{eqnarray}
\dot Q =  - Q + \Psi (Q), \label{eq6}
\end{eqnarray}
where
\begin{eqnarray}
\Psi (Q) = \sum\limits_k {kP(k)} {{{\psi _k}(Q)} \mathord{\left/
 {\vphantom {{{\psi _k}(Q)} {\left\langle k \right\rangle }}} \right.
 \kern-\nulldelimiterspace} {\left\langle k \right\rangle }}. \label{eq7}
\end{eqnarray}

\begin{figure}
\centerline{\includegraphics*[width=1.0\columnwidth]{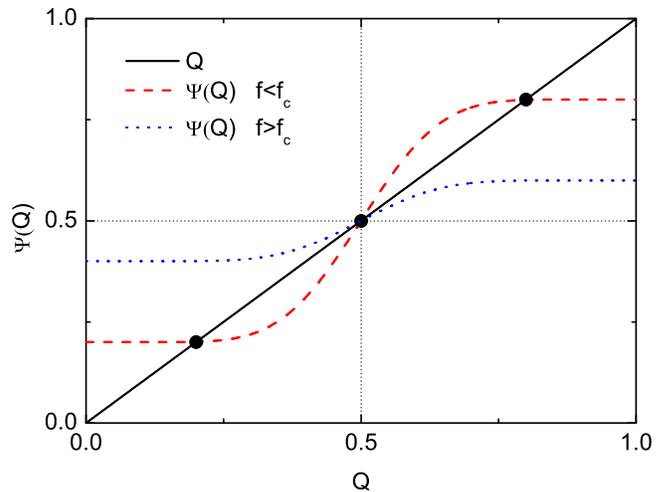}}
\caption{(color online). Graphic demonstration of steady state
solutions of $Q$. When $f$ is less than a critical value $f_c$,
there are three solutions. One is at $Q =\frac{1}{2}$ corresponding
to a disordered phase and the other two correspond to two symmetric
ordered phases. For $f>f_c$ , there is only one solution at $Q
=\frac{1}{2}$. \label{fig1}}
\end{figure}

In the steady state $\dot Q = 0$, we have $Q_s = \Psi (Q_s)$.
Fig.\ref{fig1} shows that the two typical examples of graphic
solutions of $Q_s$. One can easily find that a trivial stationary
solution, $Q_s=\frac{1}{2}$, always exists irrespective of the value
of $f$ (corresponding to a disordered phase $\left\langle {{\sigma
_i}} \right\rangle  = 0$), as ${\varphi _k}(\frac{1}{2}) =
\frac{1}{2}$ and ${\Psi}(\frac{1}{2}) = \frac{1}{2}$. However, the
other two solutions are possible if $f$ is less than a critical
value $f_c$, and they represent the existence of two ordered phases
with up-down symmetry. Therefore, the critical noise $f_c$ is
determined by the condition that the derivation of $\Psi(Q)$ with
$Q$ equals to one at $f=f_c$, i.e.,
\begin{eqnarray}
{\left. {\frac{{d\Psi (Q)}}{{dQ}}} \right|_{Q = \frac{1}{2}}} = 1.
\label{eq8}
\end{eqnarray}
To do this, we rewrite approximately Eq.(\ref{eq5}) as
\begin{eqnarray}
{\varphi _k}(Q) = {\varphi _k}(\frac{1}{2} + y) = \frac{1}{2} +
\frac{1}{2}erf\left( {y\sqrt {2k} } \right),\label{eq9}
\end{eqnarray}
where $erf(x)$ is the error function. Note that this approximation
is plausible for large values of $k$ as the binomial distribution
can be approximated by a normal distribution and the sum over $n$ in
Eq.(\ref{eq5}) can be substituted by an integral \cite{JSM2006}. The
derivation of $\Psi(Q)$ with $Q$ can be expressed analytically by
\begin{eqnarray}
{\left. {\frac{{d\Psi (Q)}}{{dQ}}} \right|_{Q = \frac{1}{2}}} &=& {\left. {\sum\limits_k {\frac{{kP(k)}}{{\left\langle k \right\rangle }}(1 - 2f)} \frac{{d\varphi (\frac{1}{2} + y)}}{{dy}}} \right|_{y = 0}}  \nonumber \\
& = &{\left. {\sum\limits_k {\frac{{kP(k)}}{{\left\langle k \right\rangle }}(1 - 2f)} \sqrt {\frac{{2k}}{\pi }} {e^{ - 2k{y^2}}}} \right|_{y = 0}} \nonumber \\
 &= &(1 - 2f)\sqrt {\frac{2}{\pi }} \frac{{\left\langle {{k^{{3
\mathord{\left/
 {\vphantom {3 2}} \right.
 \kern-\nulldelimiterspace} 2}}}} \right\rangle }}{{\left\langle k \right\rangle
 }},\label{eq10}
\end{eqnarray}
where $\left\langle {{k^n}} \right\rangle  = \sum\nolimits_k
{{k^n}P(k)}$ is the $n$th moment of degree distribution. Inserting
Eq.(\ref{eq10}) into Eq.(\ref{eq8}), we arrive at the analytical
expression of $f_c$,
\begin{eqnarray}
{f_c} = \frac{1}{2} - \frac{1}{2}\sqrt {\frac{\pi }{2}}
\frac{{\left\langle k \right\rangle }}{{\left\langle {{k^{{3
\mathord{\left/
 {\vphantom {3 2}} \right.
 \kern-\nulldelimiterspace} 2}}}} \right\rangle }}.\label{eq11}
\end{eqnarray}

\begin{figure*}
\centerline{\includegraphics*[width=2.3\columnwidth]{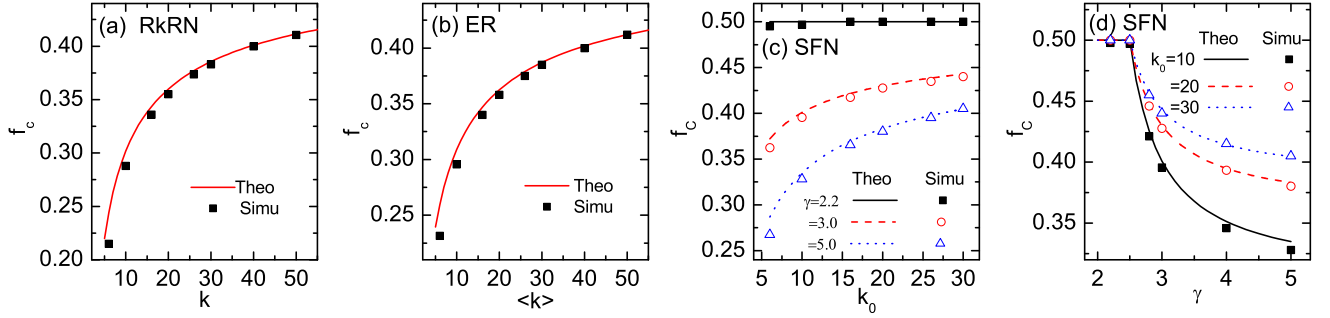}}
\caption{(color online). The critical noise $f_c$ obtained from the
theory (lines) and MC simulation (symbols) on (a) RkRN, (b) ERRN,
and (b-c) SFN.\label{fig2}}
\end{figure*}

To validate the theoretical results on $f_c$, we shall consider the
three network types: the random $k$-regular networks (RkRN) and
Erd\"os-R\'enyi random networks (ERRN), as the representations of
degree homogeneous networks, and scale-free networks (SFN) as the
representations of degree heterogeneous networks. For RkRN, each
node has the same degree $k$ and degree distribution follows the
delta-like function $P(k) = \delta (k)$, and thus Eq.(\ref{eq11})
can be reduced to
\begin{eqnarray}
f_c^{RkRN} = \frac{1}{2} - \frac{1}{2}\sqrt {\frac{\pi }{2}}
\frac{1}{{\sqrt { k  } }}.\label{eq12}
\end{eqnarray}
From Eq.(\ref{eq12}), we see that $f_c$ is a decreasing function of
$k$. For $k=N-1\rightarrow \infty$, $f_c=\frac{1}{2}$ and RkRN thus
become the globally connected networks. For ERRN, degree
distribution follows Poisson $P(k) = {\left\langle k \right\rangle
^k}{{{e^{ - \left\langle k \right\rangle }}} \mathord{\left/
{\vphantom {{{e^{ - \left\langle k \right\rangle }}} {k!}}} \right.
\kern-\nulldelimiterspace} {k!}}$ with the average degree
$\left\langle k \right\rangle$, and the theoretical value of $f_c$
for ERRN can be numerically calculated according to Eq.(\ref{eq11}).

For SFN, degree distribution follows a pow-law function $P(k) \sim
{k^{ - \gamma }}$, with degree exponent $\gamma>2$. In the
thermodynamic limit $N \to \infty$, ${\left\langle {{k^{{3
\mathord{\left/ {\vphantom {3 2}} \right.
 \kern-\nulldelimiterspace} 2}}}} \right\rangle }$ diverges for $\gamma \leq {5 \mathord{\left/
{\vphantom {5 2}} \right. \kern-\nulldelimiterspace} 2}$, such that
the critical noise becomes $f_c=\frac{1}{2}{\kern 1pt} {\kern 1pt}$
according to Eq.(\ref{eq11}). For $\gamma > {5 \mathord{\left/
{\vphantom {5 2}} \right. \kern-\nulldelimiterspace} 2}$, both
$\left\langle k \right\rangle$ and ${\left\langle {{k^{{3
\mathord{\left/ {\vphantom {3 2}} \right.
 \kern-\nulldelimiterspace} 2}}}} \right\rangle }$ are finite in the limit of $N \to \infty$, and they
are $\gamma-$dependent given by $\left\langle k \right\rangle  =
{{(\gamma  - 1){k_0}} \mathord{\left/ {\vphantom {{(\gamma  -
1){k_0}} {(\gamma  - 2}}} \right. \kern-\nulldelimiterspace}
{(\gamma  - 2}}{\kern 1pt} )$ and $\left\langle {{k^{{3
\mathord{\left/
 {\vphantom {3 2}} \right.
 \kern-\nulldelimiterspace} 2}}}}
\right\rangle  = {{(\gamma  - 1)k_0^{{3 \mathord{\left/
 {\vphantom {3 2}} \right.
 \kern-\nulldelimiterspace} 2}}} \mathord{\left/
{\vphantom {{(\gamma  - 1)k_0^{1.5}} {(\gamma  - 2.5}}} \right.
\kern-\nulldelimiterspace} {(\gamma  - {5 \mathord{\left/
 {\vphantom {3 2}} \right.
 \kern-\nulldelimiterspace} 2}}}{\kern 1pt} )$ with
$k_0$ being the minimal node degree. By the above analysis, we
immediately obtain the critical noise $f_c$ for SFN
\begin{equation}
f_c^{SFN} = \left\{
\begin{aligned}
&\frac{1}{2} - \frac{1}{2}\sqrt {\frac{\pi }{2}} \frac{{\gamma  - {5
\mathord{\left/ {\vphantom {5 2}} \right. \kern-\nulldelimiterspace}
2}}}{{\gamma  - 2}}\frac{{1{\kern 1pt} {\kern 1pt} {\kern 1pt}
{\kern 1pt} }}{{\sqrt {{k_0}} }},  & \gamma  > {5 \mathord{\left/
{\vphantom {5 2}}
\right. \kern-\nulldelimiterspace} 2}{\kern 1pt} {\kern 1pt} \\
&\frac{1}{2},  & \gamma  \le {5 \mathord{\left/ {\vphantom {5 2}}
\right. \kern-\nulldelimiterspace} 2}
\end{aligned}
\right.\label{eq13}
\end{equation}

We firstly generate the networks according to the Molloy-Reed model
\cite{RSA95000161}: each node is assigned a random number of stubs
$k$ that is drawn from a given degree distribution. Pairs of
unlinked stubs are then randomly joined. We then run the standard MC
simulation: at each MC step, each node is firstly randomly chosen
once on average and then make an attempt to flip spin with the
probability according to Eq.(\ref{eq1}).

In order to numerically obtain the $f_c$, we need to calculate the
Binder's fourth-order cumulant $U$, defined as
\begin{eqnarray}
U = 1 - \frac{{\left[ {\left\langle {{m^4}} \right\rangle }
\right]}}{{3{{\left[ {\left\langle {{m^2}} \right\rangle }
\right]}^2}}},\label{eq14}
\end{eqnarray}
where $m={{\sum\nolimits_{i = 1}^N {{\sigma _i}} } \mathord{\left/
 {\vphantom {{\sum\nolimits_{i = 1}^N {{\sigma _i}} } N}} \right.
 \kern-\nulldelimiterspace} N}$ is the average magnetization per node, $\left\langle  \cdot  \right\rangle$ denotes time averages
taken in the stationary regime, and $\left[  \cdot  \right]$ stands
for the averages over different network configurations. The critical
noise $f_c$ is estimated as the point where the curves $U\sim f$ for
different network sizes $N$ intercept each other. In our
simulations, $f_c$ is determined by five different network sizes:
$N=500$, $1000$, $2000$, $5000$ and $10000$.

For comparison, in Fig.\ref{fig2} we plot the $f_c$ obtained from
the theoretical prediction (lines) and the MC simulation (symbols),
respectively. In Fig.\ref{fig2}(a) and Fig.\ref{fig2}(b), we show
the results on RkRN and on ERRN, respectively. In
Fig.\ref{fig2}(c-d), we show the results on SFN and plot the $f_c$
as a function of $k_0$ for some fixed $\gamma$ in Fig.\ref{fig2}(c)
and of $\gamma$ for some fixed $k_0$ in Fig.\ref{fig2}(d). It is
clearly observed that for large values $k$ there are an excellent
agreements between the theory and simulation. However, for
relatively small $k$ the used approximation in Eq.(\ref{eq9}) is not
very valid, such that the discrepancy between them exists.

So far, we have obtained the analytical expression of $f_c$ and
confirmed its validity by performing MC simulations on different
networks. The expression is only valid for infinite size networks $N
\rightarrow \infty$ where the finite-size fluctuation is ignored.
For finite size networks, the fluctuation is unavoidable and the
actual phase transition never happens. However, one can define an
effective critical noise $f_c(N)$ at which the susceptibility (the
variance of an order parameter) is maximal. Obviously, $f_c(N)$ is
size-dependent and recovers $f_c$ in the limit of $N \rightarrow
\infty$. To get $f_c(N)$, we will derive the fluctuation-driven
Langevin equation for $Q$ \cite{PhysRevE.79.036110,JSM2009}:
\begin{eqnarray}
\dot Q = \mathbb{E}[\Delta Q] + \sqrt {\mathbb{V}[\Delta Q]} \xi (t)
\label{eq15}
\end{eqnarray}
where $\mathbb{E}[\Delta Q]$ and $\mathbb{V}[\Delta Q]$ are the mean
value and the variance of the variation of $Q$, respectively, and
$\xi(t)$ is a Gaussian white noise satisfying $\left\langle {\xi
(t)} \right\rangle  = 0$ and $\left\langle {\xi (t)\xi (t')}
\right\rangle  = \delta (t - t')$. For the present model,
$\mathbb{E}[\Delta Q]$ and $\mathbb{V}[\Delta Q]$ can be computed as
\begin{eqnarray}
  \mathbb{E}[\Delta Q]&=&N\sum\limits_k {P(k){q_k}[1 - {\psi _k}(Q)]\left( { - {k \over {N\left\langle k \right\rangle }}} \right)}  \nonumber\\ &&+ N\sum\limits_k {P(k)(1 - {q_k}){\psi _k}(Q)\left( { {k \over {N\left\langle k \right\rangle }}} \right)} \nonumber \\
   &=&  - Q + \Psi (Q),\label{eq16}
\end{eqnarray}
and
\begin{eqnarray}
\mathbb{V}[\Delta Q] &=& N\sum\limits_k {P(k){q_k}[1 - {\psi
_k}(Q)]} {\left( { - {k \over {N\left\langle k \right\rangle }}}
\right)^2} \nonumber \\&&+ N\sum\limits_k {P(k)(1 - {q_k}){\psi
_k}(Q){{\left( {{k \over {N\left\langle k \right\rangle }}}
\right)}^2}}  \nonumber \\ &=& \sum\limits_k {{{{k^2}P(k)} \over
{N{{\left\langle k \right\rangle }^2}}}} \left[ {q{_k}(1 - {\psi
_k}(Q)) + (1 - {q_k}){\psi _k}(Q)} \right] \nonumber \\ \label{eq17}
\end{eqnarray}
Equation (\ref{eq17}) is not yet a closed equation for $Q$ because
the diffusion term $\mathbb{V}[\Delta Q]$ involve degree-dependent
quantities $q_k$. To close it, we use the quasi-static approximation
obtained from the rate equations (\ref{eq3}) imposing $\dot
q_k\simeq0$, i.e., $q_k\simeq\psi_k(Q)$
\cite{PhysRevE.79.036110,JSM2009}. The approximation assumes that
$Q(t)$ varies much slowly with respect to the dynamics of the
microscopic degrees of freedom $q_k(t)$. By the approximation,
Eq.(17) becomes
\begin{eqnarray}
\mathbb{V}[\Delta Q] = {2 \over {N{{\left\langle k \right\rangle
}^2}}}\sum\limits_k {{k^2}P(k)} {\psi _k}(Q)(1 - {\psi _k}(Q))
\label{eq18}
\end{eqnarray}
Therefore, we obtain the fluctuation-driven Langevin equation with a
closed form,
\begin{eqnarray}
\dot Q =  - Q + \Psi (Q) + \sqrt{2D(Q)} \xi (t) \label{eq19}
\end{eqnarray}
with multiplicative noise $D(Q)={1 \over {N{{\left\langle k
\right\rangle }^2}}}\sum\nolimits_k {{k^2}P(k)} {\psi _k}(Q)(1 -
{\psi _k}(Q))$. Clearly, in the limit of $N\rightarrow\infty$, the
fluctuation term $D(Q)\rightarrow0$, and Eq.(\ref{eq19}) thus
recovers to the mean-field equation derived in Eq.(\ref{eq6}).

Furthermore, let $P(Q,t)$ denote the probability density
distribution of $Q$ at time $t$. Then, the Fokker-Planck equation of
$P(Q,t)$ corresponding to Eq.(\ref{eq19}) can be given by
\begin{eqnarray}
\frac{{\partial P(Q,t)}}{{\partial t}} =  &-& \frac{\partial
}{{\partial Q}}\left[ { - Q + \Psi (Q) + \sqrt {D(Q)D'(Q)} }
\right]P(Q,t) \nonumber \\ &+& \frac{{{\partial ^2}}}{{{\partial
^2}Q}} {D(Q)} P(Q,t) \label{eq20}
\end{eqnarray}
The stationary distribution is $P(Q)=C e^{U_{FP}(Q)}$ where $C$ is
the normalized constant and
\begin{eqnarray}
{U_{FP}}(Q) = \frac{1}{2}\ln \left[ {D(Q)} \right] - \int_{}^Q
{\frac{{ - S + \Psi (S)}}{{D(S)}}} dS \label{eq21}
\end{eqnarray}
is called the potential of the Fokker-Planck equation
\cite{Risken1992}.

The critical noise $f_c(N)$ for finite size networks is determined
using the modified susceptibility $\chi '$ defined as
\begin{eqnarray}
\chi ' = N\left[ {\left\langle {{y^2}} \right\rangle  -
{{\left\langle |y| \right\rangle }^2}} \right], \label{eq22}
\end{eqnarray}
where $y=Q-\frac{1}{2}$, $\left\langle {{|y|}} \right\rangle$ and
$\left\langle {{y^2}} \right\rangle$ are calculated by the integrals
\begin{eqnarray}
\left\langle {\left| y \right|} \right\rangle  = \int_0^1 {\left| {Q
- \frac{1}{2}} \right|P(Q)dQ} \label{eq23}
\end{eqnarray}
and
\begin{eqnarray}
\left\langle {{y^2}} \right\rangle  = \int_0^1 {{{\left( {Q -
\frac{1}{2}} \right)}^2}P(Q)dQ}, \label{eq24}
\end{eqnarray}
respectively. We expect that $\chi'$ have a peak at $f=f_c(N)$ that
diverges and $f_c(N)$ converges to $f_c$ when the network size
increases.

\begin{figure}
\centerline{\includegraphics*[width=1.0\columnwidth]{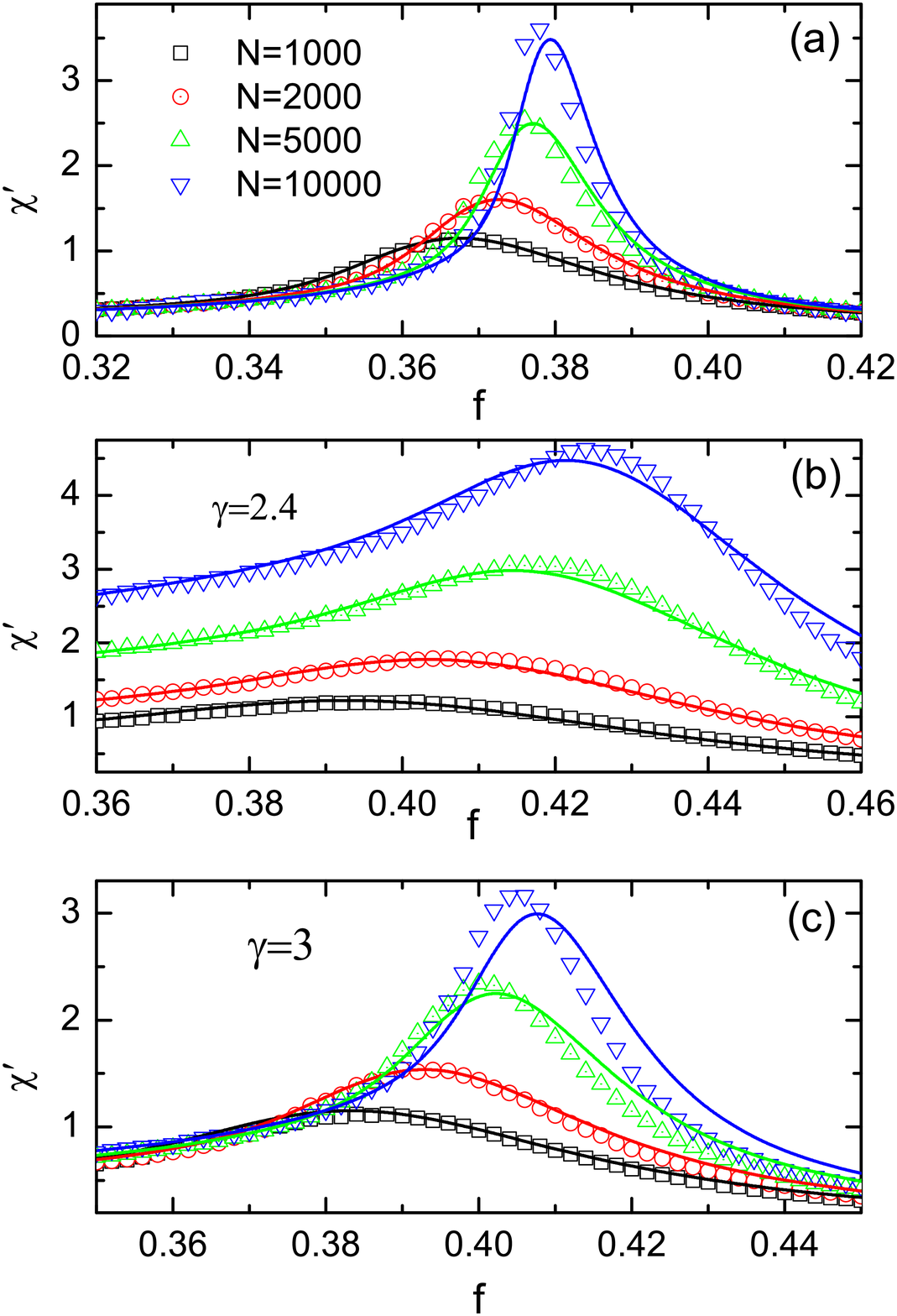}}
\caption{(color online). The susceptibility $\chi '$ as a function
of noise $f$ for some different network sizes: $N=1000, 2000, 5000,
10000$. (a) RkRN: $k=20$, (b) ERRN: $\left\langle k
\right\rangle=20$, (c) SFN: $\gamma=2.4$ and $k_0=20$, and (d)
$\gamma=3$ and $k_0=20$. Lines and symbols indicate the results from
theory and MC simulation, respectively. \label{fig3}}
\end{figure}

\begin{figure}
\centerline{\includegraphics*[width=1.0\columnwidth]{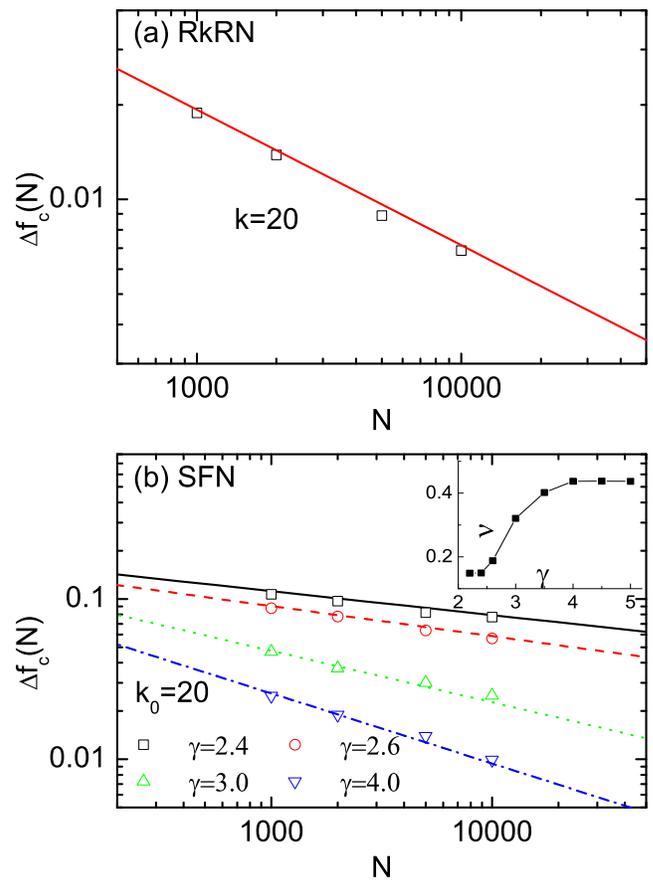}}
\caption{(color online). Log-log plot of the differences of the
effective critical noise $f_c$ for finite size networks with $f_c$,
$\Delta f(N) = {f_c} - {f_c}(N)$, with $N$ on (a) RkRN: $k=20$,
ERRN: $\left\langle k \right\rangle=20$, and (b) SFN: $k_0=20$.
$\Delta f(N)$ power law decays with $N$: $\Delta f(N) \sim {N^{ -
\nu }}$. Lines and symbols indicate the results from theory and MC
simulation, respectively. The inset of Fig.4(b) shows the exponent
$\nu$ as a function of $\gamma$. \label{fig4}}
\end{figure}

In Fig.\ref{fig3}, we show that $\chi '$ as a function of noise $f$
for some different network sizes $N$ on RkRN, ERRN, and SFN. For
comparison, the values of $\chi '$ obtained from the theoretical
calculations for Eq.(\ref{eq22}) and from MC simulations are
indicated by the lines and symbols, respectively. The theoretical
calculations can give a well prediction for simulation results. As
mentioned above, the point corresponding to the maximal $\chi '$
lies in the effective critical noise $f_c(N)$ for finite size
networks. In addition, we should note that we find that from MC
simulations the commonly used susceptibility $\chi=N\left[
{\left\langle {{m^2}} \right\rangle - {{\left\langle |m|
\right\rangle }^2}} \right]$ and our used susceptibility $\chi'$
defined in Eq.(\ref{eq22}) share the same locations where they are
maximal (results not shown here).

In Fig.\ref{fig4}, we plot the difference $\Delta f(N) = {f_c} -
{f_c}(N)$ as a function of $N$ in double logarithm coordinates. The
lines and symbols also indicate the results of theoretical
calculations from Eq.(\ref{eq22}) and MC simulations, respectively.
As pointed out by many previous studies, $\Delta f(N)$ scale with
$N$ in a power-law way: $\Delta f(N) \sim {N^{ - \nu }}$. With the
increment of $N$, $\Delta f(N)$ decreases and tends to zero in the
limit $N\rightarrow \infty$, recovering the result of
Eq.(\ref{eq11}). For RkRN and ERRN, we find that the exponents $\nu$
are independent of $k$ and $\left\langle k \right\rangle$. For SFN,
$\nu$ is also almost independent of $k_0$ but is an increasing
function of $\gamma$ (see the inset of Fig.\ref{fig4}(b)).

\section{Conclusions}
In conclusion, we have used heterogeneous mean-field theory to
derive the rate equation of an order parameter $Q$ for the MV model
defined on complex networks. By the linear stability analysis, we
have analytically obtained the critical noise $f_c$ at which the
order-disorder phase transition takes place in the limit of infinite
size networks. We find that that $f_c$ is determined by both the
first and ${3 \mathord{\left/ {\vphantom {3 2}} \right.
\kern-\nulldelimiterspace} 2}$ order moments of degree distribution
of the underlying networks. Moreover, we have incorporated the
effect of stochastic fluctuation on finite size networks via the
derivation of the Langevin equation of $Q$. By solving the
corresponding Fokker-Planck equation, we have obtained the effective
critical noise $f_c(N)$ where the susceptibility is maximal. The
results show that $f_c-f_c(N)$ power law decreases with $N$ and
reduces to zero in the limit of $N\rightarrow \infty$. To validate
the theoretical results, we have performed the extensive MC
simulations on RkRN, ERRN, and SFN. There are excellent agreement
between the theory and simulations. However, our theory does not
perform well on very sparse networks. Therefore, in the future it
will be desirable to develop high order theories (such as pair
approximation
\cite{PNAS2002,EPL2009,NJP2014,PhysRevLett.107.068701}) to obtain
more accurate estimation of the critical point of the networked MV
model.

\begin{acknowledgments}
We acknowledge supports from the National Science Foundation of
China (11205002, 61473001, 11475003, 21125313) and ``211 project" of
Anhui University (02303319-33190133).
\end{acknowledgments}

%

\end{document}